\let\Xdocument\document
\documentclass{iau}
\let\document\Xdocument

\usepackage{amsmath}
\usepackage{graphics}
\usepackage{multirow}
\usepackage{hyperref}

\begin{document}

\lefttitle{Giulia Despali}
\righttitle{Simulated lenses}

\jnlPage{1}{7}
\jnlDoiYr{2023}
\volno{381}

\doival{10.1017/xxxxx}

\aopheadtitle{Proceedings IAU Symposium}
\editors{H. Stacey, C. Grillo, A. Sonnenfeld}

\title{Gravitational lenses in hydrodynamical simulations}

\author{Giulia Despali$^{1,2}$, Felix M. Heinze$^{3}$, Claudio Mastromarino$^{4,5}$}

\affiliation{$^1$Alma Mater Studiorum - Università di Bologna, Dipartimento di Fisica e Astronomia "Augusto Righi", Via Gobetti 93/2, Bologna, Italy \\ 
$^2$ INAF - Osservatorio  di  Astrofisica  e  Scienza  dello  Spazio  di  Bologna,  via  Gobetti  93/3,  I-40129,  Bologna,  Italy \\
$^3$ Universität Heidelberg, Zentrum für Astronomie, Institut für Theoretische Astrophysik, Albert-Ueberle-Straße 2, D-69120 Heidelberg, Germany \\
$^4$ Università degli studi di Roma 'Tor Vergata', Via della Ricerca Scientifica, 1, 00133, Roma, Italy \\
$^5$ INFN-Sezione di Roma ‘Tor Vergata’, Via della Ricerca Scientifica, 1, 00133, Roma, Italy\\
}

\begin{abstract}
The gravitational lensing signal produced by a galaxy or a galaxy cluster is determined by its total matter distribution, providing us with a way to directly constrain their dark matter content. State-of-the-art numerical simulations successfully reproduce many observed properties of galaxies and can be used as a source of mock observations and predictions. Many gravitational lensing studies aim at constraining the nature of dark matter, discriminating between cold dark matter and alternative models. However, many past results are based on the comparison to simulations that did not include baryonic physics. Here we show that the presence of baryons can significantly alter the predictions: we look at the structural properties (profiles and shapes) of elliptical galaxies and at the inner density slope of subhaloes. Our results demonstrate that future simulations must model the interplay between baryons and alternative dark matter, to generate realistic predictions that could significantly modify the current constraints.

\end{abstract}

\maketitle

\section{Introduction}

One of the foundations of the concordance cosmological model is that approximately 85 percent of the universe's matter content is in the form of some yet unknown component that we can detect only through its gravitational effect: dark matter. Starting from small density perturbations in the early universe, dark matter leads the gravitational collapse on large scales, forming the cosmic-web structure of filaments along which gas flows to the potential wells of dark matter haloes, where it cools and creates stars - eventually leading to the formation of galaxies. Despite decades of experimental searches, there is yet to be a definitive answer to the most important and long-standing question in cosmology: the nature of dark matter. The favored particle candidates for cold dark matter (CDM) - e.g. axions and WIMPS, weakly interacting massive particles - have so far not been detected, motivating the study of alternative models that are allowed by particle physics assumptions and span a wide range of masses - from the eV to the GeV scale. Strong gravitational lensing is one of the most promising techniques to study the dark matter distribution of galaxies, given its ability to measure the total density distribution (both luminous and dark) in their central regions.

In the innermost regions of haloes and subhaloes, both self-interacting dark matter (SIDM) and baryons can alter the density profile in these regions. State-of-the-art hydrodynamical simulations can successfully reproduce galaxy formation and are thus an essential tool in the comparison and interpretation of lensing observations. However, constraints on alternative dark matter models have often been set by comparing observations to simulated predictions that do not include the physics of baryons. Recently, new numerical efforts have proved that it is essential to include the physics of baryons in simulations and correctly take it into account in comparison with observation. Here we present two such cases, relevant for gravitational lensing studies.

\begin{figure}
\begin{center}
 \includegraphics[width=5in]{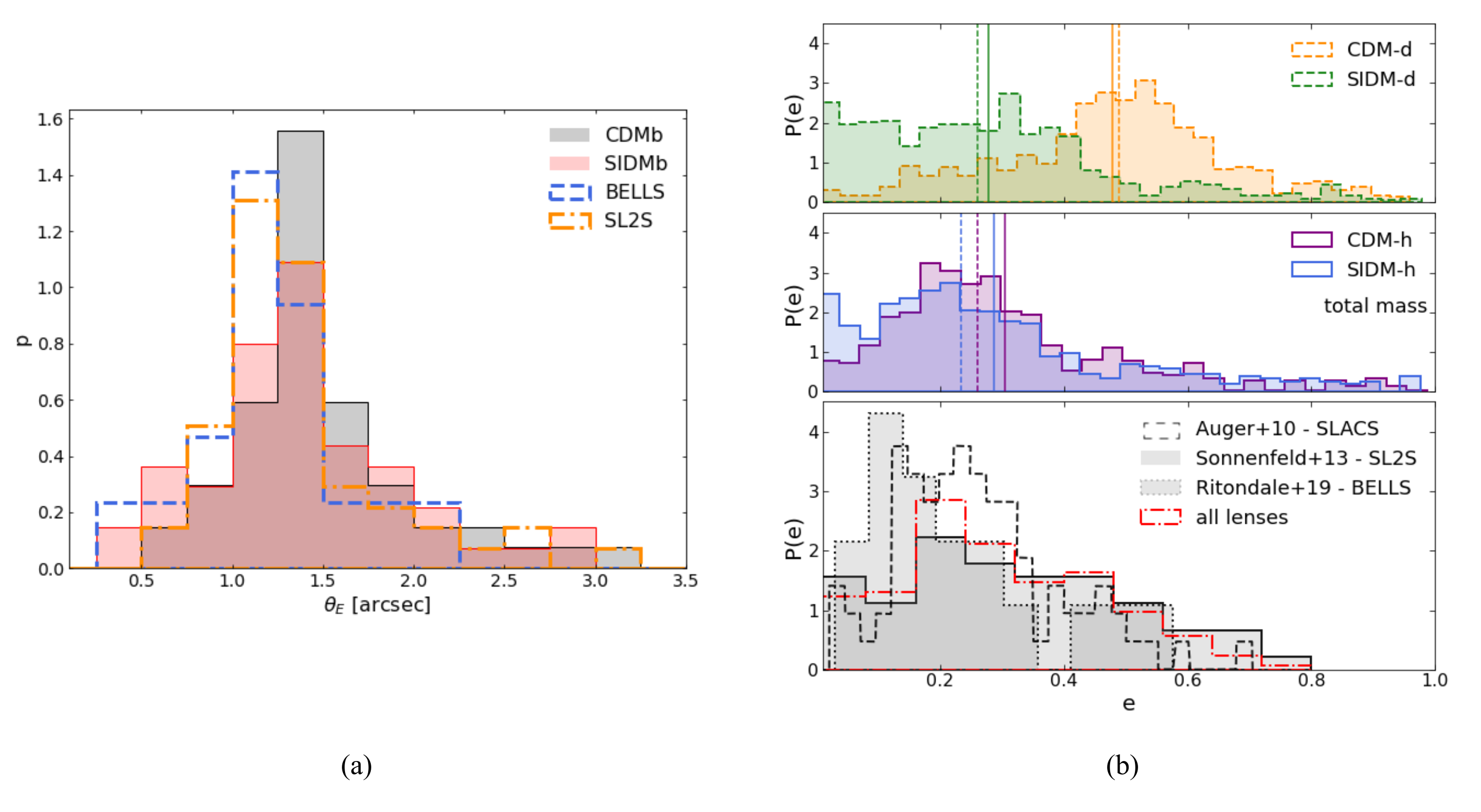} 
 \caption{Properties of gravitational lenses in hydrodynamical simulations of CDM and SIDM universes. $(a)$ Distribution of the Einstein radii $\theta_{E}$ of elliptical galaxies, measured in cold (gray) and self-interacting (red) dark matter simulations, compared to those of observed galaxy lenses from the BELLS (blue line) and SL2S (orange line) surveys. $(b)$ Distribution of projected ellipticities $e$ measured at $r=10$ kpc from the halo centre. The top and middle panel shows the results from dark-matter-only and hydrodynamical simulations, while the bottom panel reports observational measurements from gravitational lensing data sets. The figures are adapted from \cite{2023MNRAS.524.1515M} and \cite{2022MNRAS.516.4543D}.}
   \label{fig1}
\end{center}
\end{figure}

\begin{figure}
\begin{center}
 \includegraphics[width=3in]{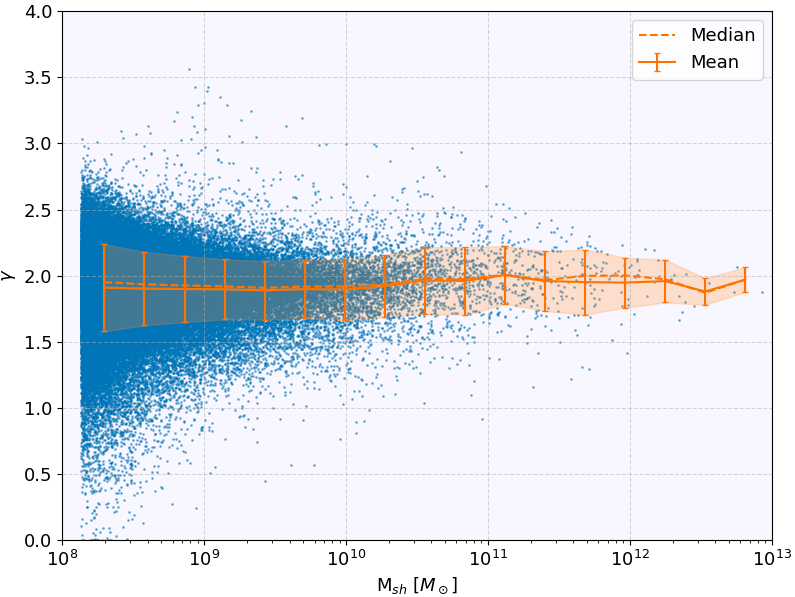} 
 \caption{Slope $\gamma$ of the total density profiles of subhaloes in the TNG50-1 hydrodynamical simulation, as a function of subhalo mass. The mean and median values for each mass bin are represented by the solid and dashed orange lines. The Figure is adapted from \cite{heinze_sub}.}
   \label{fig2}
\end{center}
\end{figure}

\section{Overview}

\subsection{Lens galaxies in SIDM universes} 

Massive systems such as elliptical galaxies and galaxy groups are the typical strong lenses at z~0.2-1: \cite{2013MNRAS.430..105P}  derived stringent constraints ($\sigma_{T}/m_{\chi} \simeq0.1cm^{2}g^{-1}$) on the self-interaction cross-section by comparing the lensing and X-ray signal in observed systems of this mass range to dark-matter-only simulations. Our recent works based on small samples of galaxies have shown that including treatment for baryonic physics can alleviate these constraints. 
\cite{2023MNRAS.524.1515M} analysed the Eagle50-SIDM simulation by \cite{2021MNRAS.501.4610R} and compared the density profiles and lensing signal of massive galaxies found in the CDM and SIDM runs. These are cosmological boxes (50 Mpc on a side) that have been re-simulated in two SIDM scenarios: $(i)$ a model with constant cross-section  $\sigma=1 cm^{2}g^{-1}$ and $(ii)$ a model where the cross-section is velocity dependent. While all dark matter haloes ($10^{12} M_{\odot}\leq M \leq 10^{14} M_{\odot}$) form a central core in the dark-matter-only runs,  the cuspy baryonic profile of the central galaxy in the hydrodynamical runs prevents the core formation and the CDM and SIDM profiles differ by less than 20 percent. Correspondingly, the Einstein radii calculated from the simulated galaxies follow the distribution shown in the left panel of Figure \ref{fig1}. Both models successfully match the observed Einstein radii from the SLACS \cite{2008ApJ...682..964B} and SL2S surveys \cite{2013ApJ...777...97S}. This demonstrates that a cross-section of $\sigma=1 cm^{2}g^{-1}$ is still compatible with lensing observations: a small effect of self-interactions at this mass scale can be expected given that the baryonic contribution to the central density is maximised for galaxies with $M\simeq10^{13}M_{\odot}$.

Similarly, the inclusion of baryons also modifies the distribution of halo shapes. In dark-matter-only simulations, CDM haloes are on average prolate and more elongated towards the centre, while self-interactions create a round inner core: the overall distribution of halo shapes is thus very different and has been used by \cite{2013MNRAS.430..105P} to put one of the most stringent constraints on the self-interaction cross-section. However, the presence of baryons can once again reduce this difference at the scale of elliptical galaxies: in \cite{2022MNRAS.516.4543D}, we ran zoom-in hydro simulations in CDM and a SIDM model with $\sigma=1 cm^{2}g^{-1}$, where galaxy formation is described by the TNG model (see \cite{2018MNRAS.473.4077P}). We calculated intrinsic and projected halo shapes, showing that both self-interactions and baryons contribute to the creation of rounder inner shapes. Overall, the distributions from the two hydro runs are compatible with each other and with the shapes measured in gravitational lensing data - see the right panel of Figure \ref{fig1}). 

\subsection{Detection of dark matter subhaloes} 

Dark matter subhaloes have been detected via the perturbations that they induce on the surface brightness of lensed arcs \cite{2010MNRAS.408.1969V}, \cite{2012Natur.481..341V}, \cite{2016ApJ...823...37H}. The strength and the shape of the perturbation depend on the density profile of the subhalo: this was originally modelled as a Pseudo-Jaffe profile (with an inner slope equal to -2), while the most recent works use the standard (and shallower) NFW profile. However, the analysis by \cite{2021MNRAS.507.1662M} has demonstrated that the subhalo concentration must be extremely high to explain the detection in the lens SDSSJ0946+1006 \cite{2010MNRAS.408.1969V}, thus questioning the predictions of Cold Dark Matter. \cite{2021MNRAS.507.1662M} compare the observational results to the structures found in the IllustrisTNG-100 simulation (\cite{2018MNRAS.473.4077P}, finding that the simulations do not contain subhaloes that can fully reproduce the data. \cite{2023arXiv230601830N} suggest that the tension could be resolved by strong self-interactions that would cause gravothermal core-collapse of some subhaloes and the formation of a very steep density profile.

In a recent work \cite{heinze_sub}, we extend the comparison to simulations to the higher resolution TNG50 simulation \cite{2019MNRAS.490.3196P}, where the spatial resolution is 288 pc (instead of 740 in TNG100). We find that the subhaloes in TNG50 have profiles that are steeper than the NFW model: the mean inner slope equals  -2, and even cuspier systems exist (see Figure \ref{fig2}). The higher spatial resolution allows to trace the density profile more accurately and find candidates that could match the observed values. This is further evidence that baryonic effects need to be taken into account and modelled down to a resolution that is appropriate to each scientific case. In a follow-up paper, we will study the lensing effect of subhaloes in detail and calculate the physical properties that fit the data.

\section{Conclusions}

In summary, once baryonic physics is properly accounted for, the lensing properties of haloes and subhaloes can be predicted more accurately both in CDM and SIDM scenarios. At the scale of massive elliptical lens galaxies, we conclude that both CDM and a SIDM model with a constant cross-section $\sigma/m_\chi =1 cm^{2}g^{-1}$ are still compatible with the most recent observations of elliptical galaxies, also consistently with the conclusions of the observational analysis by \cite{2021JCAP...05..020M}. 

Baryonic physics in CDM can also make subhaloes more compact and cuspy, increasing their lensing efficiency. It is possible that the presence of baryons can explain the extreme concentration of the subhalo detected by \cite{2010MNRAS.408.1969V}, without invoking alternative SIDM models in which subhaloes undergo gravothermal core-collapse.

It is thus evident how the inclusion of baryons in the simulated results is of fundamental importance to derive realistic predictions, both to confirm or challenge the CDM model. We compared observational results to limited samples of simulated ETGs: larger samples of simulations will help us to derive more precise predictions, avoid selection biases, and base the results on complete samples (see also \cite{2022arXiv220710638A} for a review of future prospects in SIDM models). Moreover, a more consistent analysis of the lensing data would eliminate potential sources of bias and strengthen the comparison with future simulations. Upcoming data sets and new simulations will allow the community to make significant progress in studying the nature of dark matter with astrophysical probes.

\end{document}